\documentclass[aps,showpacs,twocolumn,floatfix,amssymb,superscriptaddress]{revtex4-1}

\usepackage[british]{babel}
\usepackage[utf8]{inputenc}
\usepackage{amsmath,amssymb}
\usepackage{graphicx}
\usepackage{ae}
\usepackage{braket}
\usepackage{simplewick}
\usepackage{float}
\usepackage{simplewick}
\usepackage{color}
\usepackage[pdftex,colorlinks=true,linkcolor=blue,citecolor=blue,urlcolor=black]{hyperref}
\usepackage{import}
\usepackage{multirow}
\usepackage{comment}
\usepackage{multibib}

\begin{document}

\newcommand{\norm}[1]{\left\lVert#1\right\rVert}
\newcommand{\fpch}[1]{{\color{blue}#1}}
\newcommand{\fpcm}[1]{{\color{red}FP: #1}}

\newcommand{\red}[1]{{\color{red}{#1}}}

\def \r{{\textbf{r}}}
\def \k{{\textbf{k}}}
\def \p{{\textbf{p}}}
\def \q{{\textbf{q}}}
\def \x{{\textbf{x}}}
\def \A{{\textbf{{A}}}}
\def \a{{\textbf{{a}}}}
\def \b{{\textbf{{b}}}}
\def \c{{\textbf{{c}}}}
\def \z{{\textbf{{z}}}}
\def \0{{\textbf{{0}}}}
\def \dl{\frac{\partial}{\partial l}}
\def \P{{\boldsymbol{P}}}
\def \K{{\boldsymbol{K}}}
\def \Eb{{E_\text{b}}} 

\def\BigColSep{\setlength{\arraycolsep}{50pt}}

\title{Spectra of heavy polarons and molecules coupled to a Fermi sea}

\author{Dimitri Pimenov}\email{D.Pimenov@physik.lmu.de}
\affiliation{Arnold Sommerfeld Center for Theoretical Physics, Ludwig-Maximilians-University Munich, 80333 Munich, Germany}
\author{Moshe Goldstein}
\affiliation{Raymond and Beverly Sackler School of Physics and Astronomy, Tel Aviv University, Tel Aviv 6997801, Israel}

\begin{abstract}
We study the spectrum of an impurity coupled to a Fermi sea (e.g., minority atom in an ultracold gas, exciton in a solid) by attraction strong enough to form a molecule/trion. We introduce a diagrammatic scheme which allows treating a finite mass impurity while reproducing the Fermi edge singularity in the immobile limit. For large binding energies the spectrum is characterized by a semi-coherent repulsive polaron and an incoherent molecule-hole continuum, which is the lowest-energy feature in the single-particle spectrum. The previously predicted attractive polaron seems not to exist for strong binding.

\end{abstract}
\maketitle

 \textbf{Introduction}.---  The interaction of a single impurity with a surrounding fermionic bath is a   problem at the very heart of quantum many-body physics, which is easily formulated, and yet difficult to solve. It is characterized by a rich interplay of kinetic and interaction effects, which can strongly modify the quasiparticle (polaronic) nature of the impurity.
Controlled experimental realization and analysis of impurity physics has recently been achieved in ultracold gas setups \cite{schirotzek2009observation,
kohstall2012metastability, koschorreck2012attractive,
cetina2016ultrafast,scazza2017repulsive}, where the impurity is usually an excited hyperfine state of an atom, and the interaction strength is tunable via Feshbach resonances \cite{ketterle2008making}. An alternative are semiconductor or transition metal dichalcogenide experiments \cite{Smolka,sidler2017fermi}, where the impurity is a valence band hole or exciton, in the presence of a finite conduction band population controlled by gate voltage.

On the theory side, a major part of the literature is devoted to the computation of ground state energies following Chevy's \cite{chevy2006universal} pioneering work, which proposed an ansatz for the ground state wave function consisting of the impurity dressed by a single electron-hole pair. This ansatz works well in the polaronic regime where the impurity-bath interaction is weak, but breaks down if the formation of a molecule, or trion in the semiconductor language,  becomes favorable. This regime can be described by a complementary ansatz \cite{punk2009polaron, mora2009ground, combescot2010analytical} involving a dressed molecule. In 2D, a similar picture applies \cite{parish2011polaron,parish2013highly,
levinsen2015strongly}.

The variational energy has recently been verified using diagrammatic quantum Monte Carlo \cite{prokof2008fermi, prokof2008bold,vlietinck2013quasiparticle,PhysRevB.89.085119,kroiss2014diagrammatic,
kroiss2015diagrammatic}.
The situation is quite different for the \textit{impurity spectrum}, which is the actual quantity measured in experiments: in Monte Carlo, extracting the spectrum is difficult due to the infamous analytical continuation problem, and only few definite statements can be made \cite{goulko2016dark}.
Analytically, it has been realized that the Chevy ansatz is equivalent to the non-self-consistent $T$-matrix approach \cite{combescot2007normal}, from which spectra can easily be extracted \cite{punk2007theory,schmidt2012fermi,
efimkin2017many,schirotzek2009observation,
sidler2017fermi,scazza2017repulsive}. However, this ansatz is a priori reliable for weak coupling only. In the molecule limit, extracting the spectrum from a variational ansatz is difficult since the coefficients are not analytically known. As for the functional renormalization group \cite{schmidt2011excitation}, its accuracy is hard to assess \footnote{One-loop FRG certainly cannot capture the correct physics for heavy impurities in general \cite{kugler2018fermi}}.

Besides the interaction strength and Fermi energy, a third control parameter in the impurity problem is the impurity mass $M$. Infinitely heavy impurities are subject to Anderson orthogonality \cite{Anderson1967}, and the universal properties of the impurity spectrum in the presence of a bound state can be computed exactly from a functional determinant  \cite{combescot1971infrared,baeten2015many,
Schmidt2017, PhysRev.178.1097}. The goal of this work is to characterize the spectrum for arbitrary impurity mass, while maintaining consistency with all known limits. Building on the framework developed in our recent work \cite{pimenov2017fermi,*Pimenov2015}, we find that a rigorous expansion in the number of fermion-hole pairs reproduces the infinite mass spectrum, and obtain controlled estimates of the impurity spectrum deep in the molecular limit; in particular, we present a controlled computation of the incoherent molecular feature in the single-particle spectrum. We mostly focus on 2D for clarity, listing the modifications in 3D along the way.

\textbf{Model}.--- \
Consider a single impurity (annihilation operator $d$) immersed in a bath of fermions ($c$). In a cold atom system, the impurity can be a fermion with a quantum number different from the bath particles; in semiconducting systems, the impurity is either a valence hole plus spin degenerate conduction band bath or an exciton containing a conduction electron with a given spin, together with a bath of the opposite spin conduction electrons \footnote{for an exciton to bind two same-spin electrons, the ``trion'' angular momentum must be odd by Pauli exclusion, and this is energetically unfavourable for heavy excitons \cite{mathy2011trimers, parish2013highly}}. The usual model Hamiltonian reads:
\begin{align}
H = \sum_{\k} \left( \epsilon_\k c_\k^\dagger c_\k +  E_\k d_\k^\dagger d_\k\right)
- \frac{V_0}{\mathcal{S}} \sum_{\k,\p,\q} c^\dagger_\k c_{\k - \q} d^\dagger_\p d_{\p+\q},
\end{align}
with $\epsilon_\k = k^2/2m, E_\k = k^2/2M$. $V_0 > 0$ is the attractive contact interaction \footnote{Contact interactions provide a reasonable description of ultracold atomic systems with broad Feshbach resonances, as well as electronic systems with screened Coulomb interaction \cite{pimenov2017fermi}}, $\mathcal{S}$ the system area, and $\hbar = 1$.
Our goal is to find the single particle spectrum $\mathcal{A}(\omega)$  at zero momentum, which is proportional to the Fourier transform of the imaginary part of the retarded impurity Green function, $
D(t) = -i\theta(t) \braket{0| d_0(t) d^\dagger_0(0)|0} $, where $\ket{0}$ is the Fermi sea without impurity. We work in the real frequency formalism at zero temperature.

\begin{figure}
\centering
\includegraphics[width=.9\columnwidth]{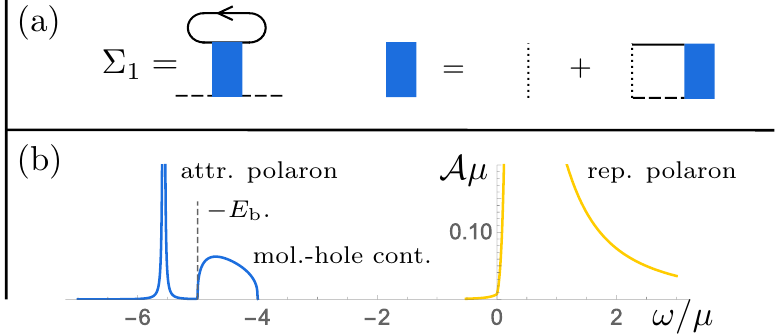}
\caption{(a) Self-energy diagrams with one hole, indicated by the arrow. Full (dashed) lines denote electron (impurity) propagators. The blue box indicates the $T$-matrix. (b) ($M = \infty$) spectrum for $\Eb  = 5\mu$ from the Chevy ansatz. Repulsive and attractive features in the spectrum are shown in different colors for clarity. For the attractive polaron, a finite width is used.}
\label{chevyfigs}
\end{figure}

\textbf{Chevy's ansatz vs.\ the Fermi edge singularity}.--- Chevy's ansatz corresponds to summation of all impurity self-energy diagrams $\Sigma_1$ with a single hole (the $T$-matrix series), shown in Fig.\ \ref{chevyfigs}(a). For infinite mass, one finds, in 2D:
\begin{align}
\label{oneholeself}
\Sigma_1(\omega) = - \int_0^\mu d\epsilon_\k  \frac{1}{\ln\left(\frac{\omega + \epsilon_\k - \mu + i0^+}{-\Eb}\right)}.
\end{align}
Here, $\omega$ is the energy measured from the impurity level, and $\mu$ is the Fermi energy. We use the standard definition of the complex logarithm with a branch cut on the negative half-axis. $-\Eb$ is the energy of the bound state of the attractive contact potential, which always exists in 2D. It is determined from the pole of the $T$-matrix. Due to this bound state, $\text{Im}[\Sigma_1](\omega)$ has a molecule continuum $\propto \theta(\omega + \Eb)$. Its width is $\mu$, representing the different energies of the hole in the Fermi sea created when the impurity binds an electron. For $\Eb \gg \mu$, inserting $\Sigma_1$ into the bare impurity Green function $D_0(\omega) = 1/(\omega + i0^+)$ leads to 3 prominent features: First, the bare pole of the impurity is shifted (``repulsive polaron''). Second, the aforementioned molecule-hole continuum is created. Third, $\text{Re}[\Sigma_1]$ gives rise to another pole below the molecule-hole continuum, the ``attractive polaron''. Between the latter two there is a spectral gap of $\simeq -0.582 \mu$ as $\Eb/\mu \rightarrow \infty$ \cite{parish2011polaron}. A typical plot is shown in Fig.~\ref{chevyfigs}(b). The 3D result is similar (see Supplemental material \cite{sm}). 
 For finite mass, the expression \eqref{oneholeself} is more complicated, but the qualitative form of the spectrum is unchanged \cite{schmidt2012fermi}.

\begin{figure}
\centering
\includegraphics[width=.9\columnwidth]{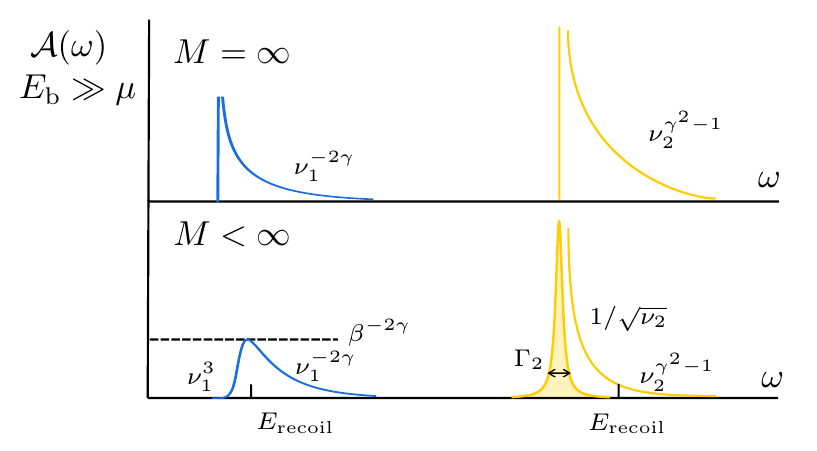}
\caption{Sketch of the spectrum for $\Eb \gg \mu$. Power laws are measured from the respective thresholds; $\Gamma_2 \simeq \mu^4/\Eb^4 \beta$ is the width of the repulsive polaron (see main text). Colors are chosen as in Fig.\ \ref{chevyfigs}.}
\label{spec_cut_fig}
\end{figure}

The ``Chevy'' spectrum for $M = \infty$ is to be contrasted with the exact result of Combescot and Nozi\`eres \cite{combescot1971infrared}, who showed that the spectrum is dominated by two divergent power laws \footnote{we neglect band bottom features discussed in Ref.\ \cite{Schmidt2017}}
$\mathcal{A}(\omega) \propto \sum_{i = 1,2}
(\omega-\omega_{\text{th}, i})^{\alpha_i} \theta(\omega - \omega_{\text{th}, i}) $. Here, $\omega_{\text{th},i}$ are the threshold energies determined from Fumi's theorem \cite{G.D.Mah2000}, and the exponents $\alpha_i$ are characterized by $\delta$, the phase shift of the bath fermions at the Fermi energy due to their scattering by the immobile impurity, $\alpha_1 = \left({\delta}/{\pi}\right)^2 - 1,  \alpha_2 = \left(1- {\delta}/{\pi}\right)^2 - 1
$. For infinite mass, the dimensionality of the problem only affects the value of $\delta$.
For $\Eb \gg \mu$ one can then approximate \cite{adhikari1986quantum,PhysRevB.41.327}
\begin{align}
\label{gammadef}
1 \gg 1-\delta/\pi \simeq \gamma \equiv \begin{cases} 1/\ln(\Eb/\mu) \ \ &\text{for}  \ \ d=2  \\ 
k_F a/\pi\ \  &\text{for}  \ \ d=3, 
\end{cases}
\end{align}
where $k_F$ is the Fermi momentum, and $a$ the 3D scattering length.
 In this limit, with exponents to leading order in $\gamma$, the spectrum looks like
\begin{align}
\label{Aexpanded}
& \mathcal{A}(\omega) \simeq  \theta(\nu_1) \nu_1^{-2\gamma} +  \theta(\nu_2)  \nu_2^{\gamma^2 -1}, \ \nu_i \equiv \omega - \omega_{\text{th,i}}.
\end{align}
A sketch is shown in Fig.\ \ref{spec_cut_fig} (upper panel). The lower (blue) feature, which starts close to $\omega = -\Eb$ and corresponds to the molecule-hole continuum, has a weak power law (close to a step). The upper feature, which can be identified with the repulsive polaron, has a strong power law spectrum (close to a delta function). Note that there is no well-defined ``attractive polaron'' in the spectrum. We claim that, for $\Eb \gg \mu$, this will persist for finite masses $M$, and thus the Chevy spectrum of Fig.\ \ref{chevyfigs}(b) is incorrect \emph{for large binding energies}.

\textbf{Method}.---
Our approach is to reproduce Eq.\  \eqref{Aexpanded} in a diagrammatic expansion in $\gamma$, which can then be generalized to finite mass. However, $\gamma$ does not directly appear in the Hamiltonian; instead, one must resort to an expansion in the \textit{number of holes}: A diagram involving $n$ holes contains $n$ integrations over filled states $\propto \mu^n$, and $\mu$ is small in units of $\Eb$. In effect, as shown below, this leads to an expansion in $\gamma$  \cite{bloom1975two,combescot2007normal,
combescot2010analytical,parish2013highly,
levinsen2015strongly}. 

\begin{figure}
\centering
\includegraphics[width=.9\columnwidth]{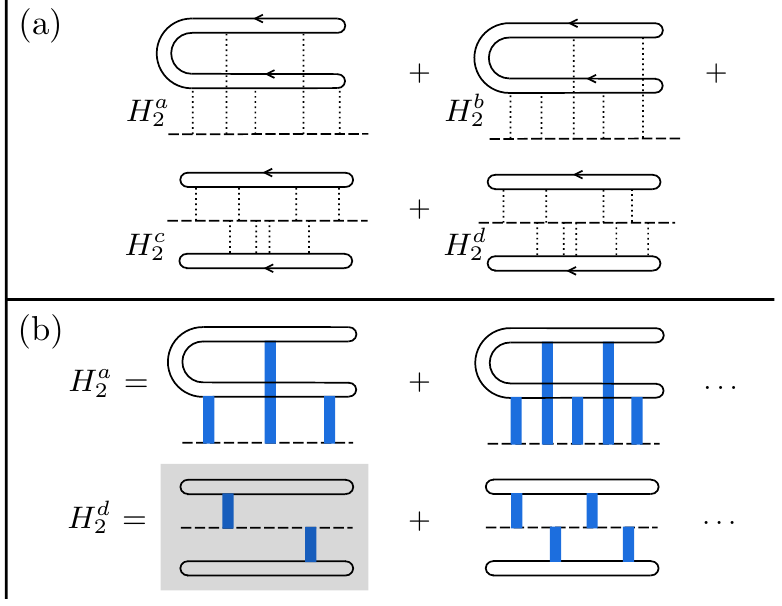}
\caption{(a) All relevant two-hole diagrams. (b)  $T$-matrix representation of diagram series $H_2^a, H_2^d$. The gray-shaded diagram is contained in the Chevy approach.}
\label{diagfig}
\end{figure}

The one-hole diagrams are already taken into account as the impurity self-energy within the Chevy approach [Fig.\ \ref{chevyfigs}(a)], and resummed with Dyson's equation. For heavy impurities, this resummation is uncontrolled. Instead, one must add up the most important (log-divergent) diagrams order by order in $\gamma$, which ultimately removes the attractive polaron from the spectrum. Thus, we reattach the impurity lines to $\Sigma_1$, defining $H_1(\omega) = D_0(\omega)^2 \Sigma_1(\omega)$. Of course, $H_1$ only represents the first order process: the impurity can interact with an arbitrary number of electrons, creating electron-hole excitations in the Fermi sea. The processes involving two holes are represented in Fig.\ \ref{diagfig}(a).
Here, the interaction lines can be drawn arbitrary often in any order, as long as the structure of the diagrams is preserved: e.g., in diagram $H_2^a$ the first and last interaction lines should connect to the lower part of the ``horseshoe'', and to the upper loop in diagram $H_2^c$. These diagrams can also be redrawn with $T$-matrix blocks, as exemplary shown in Fig.\ \ref{diagfig}(b). 
Let us point out that we never expand in the number of $T$-matrices, but always resum diagrams with an infinite number of $T$-matrices at two-hole level. We note that the contribution of the two-hole diagrams to the ground state energy is much less significant \cite{combescot2008normal, combescot2010analytical}.

\textbf{Results: The molecule/attractive polaron spectrum}.---
For infinite mass, since the impurity is dispersionless, the evaluation of all two-hole diagrams is possible. Following Ref.\  \cite{pimenov2017fermi}, one can either work in the time or frequency domain, employing different approximations \cite{sm}. In particular, for small $ \omega + E_B$,   close to the molecular threshold, we find \footnote{the computation is restricted to a parametrically large window of not too small frequencies, $\mu^2/\Eb \ll \omega + \Eb \ll \mu$ \cite{sm}}
\begin{align}
\label{H1+H2}
& H_1(\omega) + H_2(\omega) \simeq \\ & \frac{1}{\Eb}\left(\ln\left(\frac{\omega + \Eb +i0^+}{-\mu}\right)  - \gamma \ln^2\left(\frac{\omega + \Eb + i0^+}{-\mu}\right)\right), \notag
\end{align}
where $H_2 = \sum_{i} H_2^i$. The term $\propto \ln^2$ in Eq.\ \eqref{H1+H2} arises solely from diagram $H_2^a$. Curiosly, the contribution of $H_2^d$ is subleading, while the contribution from diagrams $H_2^b, H_2^c$ effectively shifts the bound state energy as $\Eb \rightarrow \Eb + \mu(1-\gamma)$  in 2D, or $\Eb + \mu(1-2\gamma/3)$ in 3D, in agreement with Fumi's theorem \cite{G.D.Mah2000} to leading order in $\gamma$.  Redefining $\nu_1$ to include these shifts, we find a contribution to the spectrum
\begin{align}
\label{A1pert}
\mathcal{A}_1(\nu_1) \simeq \frac{\theta(\nu_1)}{\Eb}\left(1- 2\gamma\ln[\nu_1/\mu]\right),
\end{align}
in agreement with Eq.\ (\ref{Aexpanded}) when expanded in $\gamma$. Note that this expansion has the same form as the perturbative expansion of the polarization in the standard Fermi-edge singularity case \cite{PhysRev.153.882,roulet1969singularities,
nozieres1969singularities}. This was to be expected, as in the limit $\Eb \rightarrow \infty$ we can formally regard the diagrams $H_{1,2}$ as polarization diagrams containing a molecule and a bath fermion, with an effective molecule-bath interaction $\gamma$. We expect higher order leading logarithmic (parquet) contributions to arise in a similar fashion from diagrams containing a larger number of holes.

Let us now address the modification of the molecule-hole feature for large but finite impurity mass $M$. The general strategy is to reevaluate the frequency-domain diagrams of Fig.\ \ref{diagfig}(b) for finite mass \cite{sm}, and trace the modification of the logarithmic singularities  \cite{gavoret1969optical,PhysRevB.35.7551, Nozi`eres1994, rosch1995heavy,rosch1999quantum,
pimenov2017fermi}. Our results hold to leading order in the mass ratio $\beta = m/M$ only, but we expect them to be qualitatively correct all the way up to $\beta \simeq 1$. First, introducing a finite mass shifts the binding energy, $\Eb \rightarrow \tilde \Eb$, but we will not compute those shifts in detail, limiting ourselves to the form of the spectrum. In terms of $ \nu_1 = \omega + \tilde \Eb$, the real part of the logarithmic singularities is modified as
$\ln \left(\max[{\nu_1 - \beta\mu},\gamma^2 \beta\mu]/\mu\right)$,
again reminiscent of the Fermi edge singularity case \cite{gavoret1969optical}. In contrast to $M = \infty$, the logarithmic singularities for finite mass are peaked at $ \nu_1 =\beta\mu$ (``direct threshold'' \cite{gavoret1969optical}). This is simply understood: when an incoming zero momentum impurity binds an electron and leaves a low-energy hole behind (as described by diagram $H_1$, yielding the first order logarithm), the resulting molecule must have a momentum $\simeq k_F$ by momentum conservation. Since the molecule is now mobile, with mass $M_+ = M + m$, one must pay its recoil energy $E_{\text{recoil}} \simeq \beta\mu$, which shifts the maximum of the logarithms to $\nu_1 =  \beta\mu$. Subsequently, the so created molecule can decay into a zero-momentum state, by exciting an electron-hole pair. The rate of this indirect process is $\Gamma_1 = \gamma^2 \beta \mu$, leading to a cutoff of the logarithmic singularities. Mathematically, this cutoff arises from diagram $H_2^c$, which can be interpreted as a molecule self-energy diagram with imaginary part $\Gamma_1$.
For large frequencies, $\nu_1 \gg E_{\text{recoil}}$, one recovers the infinite mass behavior $\propto \nu_1^{-2\gamma}$.

Apart from cutting off the singularity, the decay of the molecule leads to a shift of the threshold from the direct to the ``indirect'' one at $ \nu_1 = 0$, which corresponds to creation of zero momentum molecules. Near the indirect threshold, the spectrum starts continuously, with a power law $\propto  \nu_1^3$ in 2D and $\propto \nu_1^{7/2}$ in 3D. This behavior is obtained by computing the imaginary parts of diagrams $H_2^{a,c}$, which yield the leading contributions in $\gamma$ via standard phase space estimation \cite{pimenov2017fermi,sm}. For a spinless Fermi sea, the two contributions cancel; however, even in this case we expect that the power law behavior is robust, since it is (a) determined from a generic phase space estimate and (b) there may well be processes involving 3 holes that yield the same behavior. Exponentiating the logarithms \cite{pimenov2017fermi}, one finds the spectrum near both tresholds to be
\begin{align}
\label{cutoffFermiedge}
\mathcal{A}_1(\nu_1) \simeq \frac{1}{\Eb} \left(\frac{\sqrt{( \nu_1 - \beta\mu)^2 + (\gamma^2\beta\mu)^2}}{\mu}\right)^{-2\gamma} \!\theta(\nu_1) f_1( \nu_1),
\end{align}
where $f_1(\nu_1)$ smoothly interpolates between $f_1(\nu_1) \simeq \gamma^2 (\nu_1/\beta\mu)^{3}$  in 2D and  $f_1(\nu_1) \simeq \gamma^2 (\nu_1/\beta\mu)^{7/2}$  in 3D, for $\nu_1 \ll \beta\mu $, and $f_1(\nu_1) \simeq \pi \text{ for }  \nu_1 \gtrsim \beta\mu$. A typical plot of the resulting spectrum is shown in Fig.\ \ref{spec_cut_fig} (blue feature in lower panel). Let us reiterate the main point: the ground state signal in the spectrum is  purely incoherent, with maximum $\propto (\beta)^{-2\gamma}$ \footnote{the coefficient in front of $\beta$ cannot be determined exactly with logarithmic accuracy, but is between $\gamma^2$ and $1$ \cite{gavoret1969optical, pimenov2017fermi}}; there is no polaronic delta peak.

\textbf{Results: The repulsive polaron spectrum}.---
We now discuss the repulsive polaron, which is already predicted by the Chevy ansatz \cite{cui2010stability, massignan2011repulsive, schmidt2011excitation, ngampruetikorn2012repulsive,parish2013highly,
scazza2017repulsive}. In fact, for $\Eb \gg \mu$, the repulsive polaron contains most of the spectral weight, $\sim 1 - \mu/\Eb$, as  seen in Fig.\ \ref{chevyfigs}(b): As $\Eb/\mu\rightarrow \infty$, the repulsive polaron is essentially a spectral probe of the  impurity without Fermi sea, with unit weight. For infinite mass, the asymptotic form of the repulsive polaron is given by the second term in Eq.\ \eqref{Aexpanded}, with $\omega_{\text{th},2} \simeq \gamma \mu$ in 2D and $\omega_{\text{th},2} \simeq \frac{2}{3}\gamma \mu$ in 3D.   To leading order in $\gamma$, $\mathcal{A}_2(\nu_2) \simeq \gamma^2 \theta(\nu_2)/\nu_2$, which reduces to a delta-function as $\gamma \rightarrow 0$. This leading order term can already be obtained from the first order diagram $H_1$ for small positive frequencies. One can also reproduce the full power law singularity in a linked cluster approach, formally exponentiating $H_1$. Extending the latter approach to finite mass, one finds a delta peak with weight $\beta^{\gamma^2}$, on top of an incoherent background $\propto 1/\sqrt{\nu_2}$ for $\nu_2 \ll \beta\mu$, similar to the results of Rosch and Kopp \cite{rosch1995heavy}. In 3D, the incoherent part is approximately constant. For much larger frequencies $\nu_2 \gtrsim \beta\mu$, one recovers the infinite mass behavior $\propto \nu_2^{\gamma^2 -1}$ \cite{sm}. 

Thus, in a first approximation, the repulsive polaron is a delta-peak plus incoherent background. However, for finite mass, the delta-peak may be broadened due to decay into the low-laying molecule-hole continuum, resulting in a finite width $\Gamma_2$. This width can be estimated by computing the self-energy part of the diagrams $H_2$ (called $\Sigma_2$) at the repulsive polaron threshold $\nu_2 = 0$. Note that, for infinite mass the problem becomes single-particle \cite{PhysRev.178.1097}, forbidding such a transition; this behavior is reproduced by our calculations.
Unfortunately, for finite mass a complete evaluation of $\text{Im}[\Sigma_2(\nu = 0)]$ is out of reach. A simple estimate can be obtained from a Golden-Rule type expansion of $\Sigma_2$ in $T$-matrices \cite{sm}, similar to Ref.\ \cite{bruun2010decay};  we find, in 2D, $\Gamma_2 \sim \gamma^2 \beta \frac{\mu^4}{\Eb^3}$; in 3D, $\Gamma_2$ should still be small in $\mu/\Eb$, but the scaling could be different. Putting everything together, an approximate expression for the repulsive polaron spectrum is given by
\begin{align}
\label{A2main}
\mathcal{A}_2(\nu_2) \simeq (\beta)^{\gamma^2} \!  \!  \frac{\Gamma_2}{\nu_2^2 + (\tfrac{1}{2}\Gamma_2)^2} + f_2(\nu_2),
\end{align}
where $f_2(\nu_2)$ interpolates between the limits $f_2(\nu_2) \simeq \gamma^2/{\sqrt{\beta\mu \nu_2} } \text{ in 2D and } f_2(\nu_2) \simeq \gamma^2/(\beta\mu) \text{ in 3D}, \text{for } \nu_2 \ll \beta\mu, $ and $f_2 \simeq 1/\mu (\nu_2/\mu)^{\gamma^2 -1} \text{ for } \nu_2 \gtrsim \beta\mu$. A sketch is shown in Fig.\ \ref{spec_cut_fig} (yellow feature in the lower panel).

\begin{figure}
\centering
\includegraphics[width=\columnwidth]{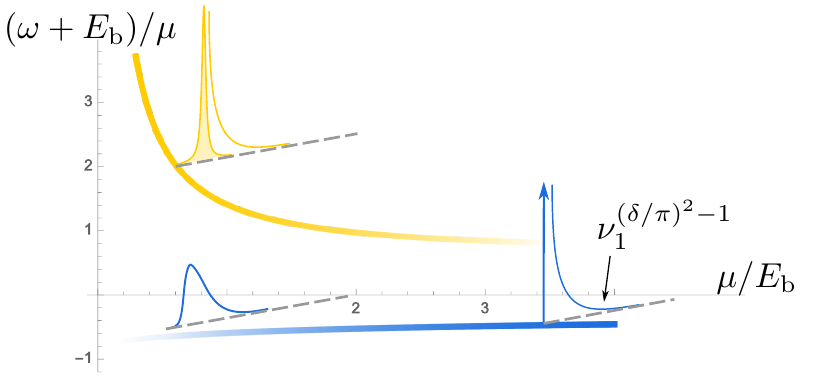}
\caption{Sketch of the full 2D spectrum for general values of $\mu/\Eb$. Thick lines indicate the position of the thresholds determined from Fumi's theorem.}
\label{fullsketch}
\end{figure}

\textbf{Discussion}.---
So far, we only discussed the spectrum in the molecular limit $\Eb \gg \mu$. In the opposite limit, the influence of the bound state should be neglible. The spectrum of a heavy impurity without a bound state was computed in \cite{rosch1995heavy}, and we expect the same result here: a single feature of a form similar to the repulsive polaron described above, but with a delta-peak that is not broadened, and with singularity exponents controlled by $\delta \ll 1$ for $\mu \gg \Eb$. Both known limits (in 2D) are sketched in Fig.\ \ref{fullsketch}, along with the thresholds as determined from Fumi's theorem, which should be approximately correct for large masses. Note that if we follow the lower spectral feature, we see a ``molecule-to-polaron transition'', in the sense that, for $\mu \ll \Eb$, the single particle spectrum is fully incoherent, but fully coherent in the opposite limit. However, the details of this transition/crossover \cite{edwards2013smooth} remain to be explored. In particular, it would be interesting to analyze this in 3D, where a bound state only forms at a certain strength of the interparticle attraction.

Let us also comment on the connection to quantum Monte-Carlo and experiments. A major difference is that the Monte-Carlo works extract the molecule solely from a pole in the two-particle propagator. The latter can be obtained from our recent work \cite{pimenov2017fermi}, and we found essentially opposite behavior to the one presented here; e.g., for $\Eb \gg \mu$, there is a sharp feature related to the molecule, and a broad continuum at larger energies. However, in the present work we have argued that the molecule emerges as an incoherent ground-state feature in the single-particle propagator as well. This seems to be in agreement with the ultracold gas experiments in both 3D  \cite{schirotzek2009observation,
kohstall2012metastability,  cetina2016ultrafast,scazza2017repulsive} and 2D \cite{koschorreck2012attractive}, while the results of the 2D TMD experiment are somewhat less clear \cite{sidler2017fermi}. The incoherent molecule feature was not seen in the ``polaron-spectra'' of the recent Monte-Carlo work \cite{goulko2016dark}, which could be attributed to problems with analytical continuation. Finally, let us note that most Monte-Carlo works, in 3D \cite{prokof2008fermi, prokof2008bold,vlietinck2013quasiparticle, kroiss2015diagrammatic, goulko2016dark} and 2D \cite{PhysRevB.89.085119, kroiss2014diagrammatic},  deal with the (almost) equal mass case, while in the experiment also heavily mass-imbalanced $^6\text{Li}-^{40}\!\text{K}$ mixtures are used. Anyway, we do not expect significant changes in the spectra for equal masses, except for the disappearance of the orthogonality power laws beyond $E_{\text{recoil}}$.

\textbf{Conclusion}.---
We presented a controlled computation of polaron spectra, providing the connection to the infinite mass limit. We found that, for large binding, the attractive polaron and molecule-hole continuum merge into a single incoherent feature, and also gave a detailed description of the repulsive polaron spectrum. Our work paves the way towards the study of many impurity physics, including the effective interaction between impurities, molecular condensate vs.\ polaron Fermi gas, etc. \cite{radzihovsky2010imbalanced,hu2018attractive}.

\textbf{Acknowledgment}.--- The authors acknowledge very helpful discussions with J.\ von Delft, L.\ I.\ Glazman,  O.\ Goulko, S. Huber,  A.\ Imamo\u{g}lu, L.\ Pollet, N.\ Prokof'ev, M.\ Punk, and R.\ Schmidt. D.\ P.\ and M.\ G.\ were supported by the German Israeli Foundation (Grant No. I-1259-303.10).
In addition, D.\ P.\ was supported by
the German Excellence Initiative via the Nanosystems
Initiative Munich (NIM), and M.\ G.\ was supported by the Israel Science Foundation (Grant No.\ 227/15), the US-Israel Binational Science Foundation (Grant No. 2014262), and the Israel Ministry of Science and Technology (Contract No.\ 3-12419).

\bibliographystyle{apsrev4-1}
\bibliography{attractive_polarons}

\clearpage  \newpage

\renewcommand{\thesection}{S.\Alph{section}}
\setcounter{figure}{0}
\renewcommand{\thefigure}{S\arabic{figure}}
\setcounter{equation}{0}
\renewcommand{\theequation}{S\arabic{equation}}

\section*{Supplementary Material for ``Spectra of heavy polarons and molecules coupled to a Fermi sea''}

In this supplement, we present the detailed  derivation of our results. In Sec.\ \ref{Fumisec} we recapitulate Fumi's theorem and the determination of the phase shift $\delta$. The evaluation of the one-hole diagram $H_1$ in the infinite mass limit is presented in Sec.\ \ref{H1app} for 2D, and in Sec.\ \ref{H1app3D} for 3D. Focusing on 2D, the two-hole diagrams $H_2$ are computed in the time domain in Sec.\ \ref{timedom_supp} and in the frequency domain in Sec.\ \ref{freq-domain}. Then, in Sec.\ \ref{mass-app}, the molecule-hole continuum is determined for finite mass. Finally, the linked cluster approach is used to compute the repulsive polaron for infinite mass in Sec.\ \ref{linked cluster}, and for finite mass in Sec.\ \ref{modlinkc}.

\section{Determination of the phase shifts and Fumi's theorem}
\label{Fumisec}

\begin{figure}
\centering
\includegraphics[width=\columnwidth]{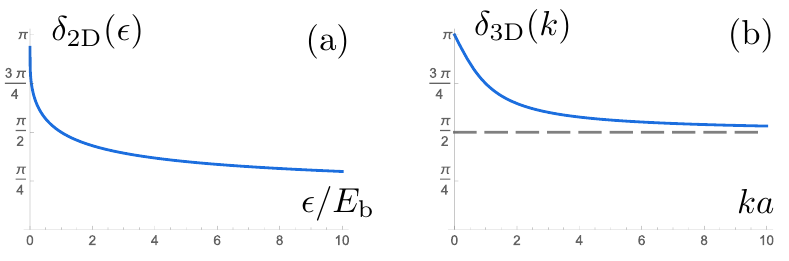}\caption{Scattering phase shifts (a) 2D phase shift (c) 3D phase shift. }
\label{deltaplot}
\end{figure}

The universal properties of immobile impurities are characterized by the energy dependent scattering phase shift of the two-particle problem, $\delta(\epsilon)$. In particular, the power law exponents discussed in the main text are determined by the phase shift at the Fermi energy, $\delta(\mu)$. For a zero-range interaction potential in 2D, $\delta$ is given by \cite{adhikari1986quantum,PhysRevB.41.327}
\begin{align}
\label{phaseshift_eq}
\delta_{\text{2D}}(\epsilon) =  \cot^{-1}\left(\ln\left[\frac{\epsilon}{\Eb}\right]/\pi\right),
\end{align}
where $\Eb$ is the 2D binding energy (chosen as positive, i.e., the bound state occurs at $-\Eb$). In 3D, provided the potential is strong enough to form a bound state, the phase shift is usually presented as \cite{PhysRevB.41.327}
\begin{align}
\label{phaseshift3d_eq}
\delta_{\text{3D}}(k) = \pi + \arctan(-ka), \quad k = \sqrt{2 m \epsilon}, 
\end{align}
with the 3D scattering length $a$. Note that the reduced mass $m_r$ equals $m$ for immobile impurities. $\delta_{\text{2D}}$ and $\delta_{\text{3D}}$ are plotted in Fig.\ \ref{deltaplot}.
In terms of $\delta$, the lower threshold of the spectrum $\omega_{\text{th},1}$ is obtained from Fumi's theorem (see, e.g., \cite{G.D.Mah2000}):
\begin{align}
\label{Fumithreshold}
\omega_{\text{th}, 1} = - \Eb - \int_0^\mu \frac{d\epsilon}{\pi} \  \delta(\epsilon), 
\end{align}
with $\Eb = 1/(2m a^2)$ for $M = \infty$ in $d = 3$.
The second threshold is reached by removing the Fermi sea electron from the bound state and putting it on top of the Fermi sea, thus
$\omega_{\text{th},2} = \omega_{\text{th,1}} + \Eb + \mu$. The two thresholds are plotted in Fig.\ \ref{fullsketch} of the main text.

\section{$M = \infty$: Evaluation of $H_1$ in 2D} \label{H1app}

The $T$-matrix  corresponds to the sum of all ``ladder''-diagrams for the two-particle vertex and is pictorially defined in Fig.\ \ref{chevyfigs}(a) of the main text. It is solely a function of total energy-momentum $(\omega + \epsilon, \k)$. In the evaluation of the $T$-matrix and all further diagrams, the summation over internal frequencies is trivial: the bare propagator of the single impurity is purely retarded (since there is no ``impurity Fermi sea''), which effectively sets all internal frequencies on-shell and restricts the momenta of electrons propagating forward (backward) in time to be above (below) $k_F$. With finite impurity masses, in 2D the $T$-matrix is therefore given by:
\begin{align} \notag
&T(\omega + \epsilon, \k) = \\ & \left(-1/V_0 -
\int  d\p  \frac{1}{\omega + \epsilon - \epsilon_\p  - E_{\k - \p} + i0^+}\right)^{-1}.
\label{Tmatrixdef}
\end{align}
Here and henceforth, we use the convention
\begin{align}
\label{intconvention}
\int  d\k  = \int_{k< k_F} \frac{d^2 p}{(2\pi)^2}, \   \int d\p  d\q   = \int_{k_F < p,q< p_\xi} \frac{d^2 p}{(2\pi)^2} \frac{d^2 q}{(2\pi)^2},
\end{align}
where $p_\xi = \sqrt{2m \xi}$, and $\xi$ is a UV cutoff. The pole of Eq.\ \eqref{Tmatrixdef} for $\k, k_F \rightarrow 0$ defines the vacuum binding energy $-\Eb$ \cite{parish2011polaron}. One can also define a 2D scattering length $a_{\text{2D}} = \sqrt{2 m_r \Eb}$ with reduced mass $m_r$, but we will not use this quantity further.

For $M =  \infty$, the impurity is dispersionless, $E = 0$, and Eq.\ \eqref{Tmatrixdef} reduces to
\begin{align}
\label{Tinfmass}
T(\omega + \epsilon) = - \frac{1}{\rho} \frac{1}{\ln\left(\frac{\omega + \epsilon - \mu + i0^+}{-\Eb}\right)},
\end{align}
with the density of states $\rho = m/2\pi$. $\Eb$ is  given by $\xi e^{-1/(\rho V_0)}$. From Eq.\ \eqref{Tinfmass}, the one-hole self-energy is  obtained by closing the electron loop. This yields Eq.\ \eqref{oneholeself} of the main text:
\begin{align}
\label{Sigmaapp}
\Sigma_1(\omega) = - \int_0^\mu d\epsilon_\k  \frac{1}{\ln\left(\frac{\omega + \epsilon_\k - \mu  + i0^+}{-\Eb}\right)}.
\end{align}
The one-hole diagrams $H_1$ are obtained by reattaching the impurity lines, $H_1(\omega) = \Sigma_1(\omega) D_0(\omega)^2$.

\subsection*{Molecule-hole feature}

To find the contribution of $H_1$ to the molecule feature, we expand Eq.\  \eqref{Sigmaapp} around $\omega = -\Eb$. Thus
\begin{align}
\label{H12D}
&H_1(\omega) \simeq \frac{1}{\Eb} \ln\left(\frac{\omega + \Eb + i0^+}{\omega + \Eb - \mu}\right) \\ & \simeq \frac{1}{\Eb} \ln\left(\frac{\omega + \Eb + i0^+}{ - \mu}\right) \ \text{for} \ |\omega + \Eb| \ll \mu, \notag
\end{align}
yielding the first logarithm in Eq.\ \eqref{H1+H2}.

\subsection*{Repulsive polaron}
\label{H1reppolaron}
For the repulsive polaron, we need to evaluate Eq. \eqref{Sigmaapp} for $\omega \gtrsim 0$. A useful formula is
\begin{align}
\int dx \frac{x^n}{\ln(x)} = \begin{cases} \frac{x^n}{n+1} \frac{1}{\ln x} + \mathcal{O}\left(\frac{x^n}{\ln^2(x)}\right),  \quad &n \neq -1 \\
\ln(\ln (x)) \quad &n = -1
\end{cases}.
\end{align}
Schematically, this formula implies that the $1/\log$ terms can be pulled out from integrals with ``logarithmic accuracy'' (l.a.). As a result, we find
\begin{align}
\label{sigmaexp}
\Sigma_1(\omega) \simeq - \frac{\mu}{\ln\left(\frac{\mu}{\Eb}\right)} +   \frac{\omega\cdot \ln\left(\frac{|\omega|}{\mu}\right)}{\ln\left(\frac{\mu}{\Eb}\right) \ln\left(\frac{|\omega|}{\Eb}\right)} - i \frac{\pi \omega \theta(\omega)}{\ln^2\left(\frac{|\omega|}{\Eb}\right)}.
\end{align}
Upon resummation, the first term in Eq.\ \eqref{sigmaexp} shifts the repulsive polaron threshold to $\omega = \gamma \mu \simeq \omega_{\text{th},2}$, with $\gamma$ as defined in Eq.\ \eqref{gammadef} of the main text. To interpret the other terms, we restrict ourselves to a parametrically large window of frequencies $\mu^2/\Eb \ll \omega \ll \mu$, which allows the simplification $\ln(\omega/\Eb) = \ln(\omega/\mu \cdot \mu/\Eb) \simeq \ln(\mu/\Eb)$ with l.a. This restriction is specific to our diagrammatic approach, and is not required in the infinite mass treatment \cite{combescot1971infrared}. Thus, we believe that our results hold down all the way to $\omega \rightarrow 0$. Taking into account the threshold shift, i.e.\ shifting to $\nu_2 = \omega - \omega_{\text{th}, 2}$, and reattaching the impurity lines yields:
\begin{align}
\label{H1nu2}
H_1(\nu_2) \simeq \frac{\gamma^2}{\nu_2} \ln\left(\frac{\nu_2 + i0^+}{-\mu}\right).
\end{align}
Taking the imaginary part leads to a spectrum $\mathcal{A}_2(\nu_2) \simeq \gamma^2 \theta(\nu_2)/\nu_2$ given in the main text.

\section{$M = \infty$: Evaluation of $H_1$ in 3D}
\label{H1app3D}

To substantiate our claim that our results apply to 3D in analogous fashion, here we present the evaluation of $H_1$ in 3D. We start from the infinite mass $T$-matrix analogous to Eq.\ \eqref{Tmatrixdef}: 
\begin{align}
T(\Omega) =  \left(-1/V_0 -
\int  d^3\p  \frac{1}{\Omega - \epsilon_\p  + i0^+}\right)^{-1}\!, \quad  \Omega = \omega + \epsilon.  
\label{Tmatrixdef3D}
\end{align}
The 3D integrals follow the convention of Eq.\ \eqref{intconvention} adapted to 3D (in this section only). To regularize the $T$-matrix, we apply the Lippmann-Schwinger equation \cite{ketterle2008making,combescot2007normal}
\begin{align}
\frac{1}{-V_0} = \frac{m}{2\pi a} - \int_{>0} d^3\p \frac{1}{\epsilon_\p}, 
\end{align} 
where the integral ranges over all momenta $0<p<p_\xi$. As a result, 
\begin{align}
\label{Tregularized}
&T(\Omega) =  \\ &\left(\frac{m}{2\pi a}  - \left[ \int d^3\p \frac{1}{\Omega - \epsilon_\p + i0^+} - \int_>\!d^3\p \frac{1}{-\epsilon_\p} \right] \right)^{-1}\!. \notag
\end{align}
After some straightforward algebra (see, e.g., \cite{gradshteyn2014table, Punk2010}), $T$ can be rewritten as 
\begin{align}
&T(\Omega) = \left(\frac{m}{2\pi a} - c_1 R(\Omega)\right)^{-1}, \ c_1 =  \frac{m^{3/2}}{\sqrt{2}\pi^2}, \\ &
R(\Omega) = \theta(\Omega) \left(2\sqrt{\mu} + \sqrt{\Omega} \ln \left|\frac{\sqrt{\mu} - \sqrt{\Omega}}{\sqrt{\mu} + \sqrt{\Omega}} \right|\right) +  \\ \notag &\theta(-\Omega) \left( \pi\sqrt{-\Omega} + 2\sqrt{\mu} - 2\sqrt{-\Omega} \arctan\left(\frac{\mu}{\sqrt{-\Omega}}\right) \right) \\ & \notag -i\pi \sqrt{\Omega}\cdot  \theta(\Omega - \mu).
\end{align}
In the vacuum limit $\mu \rightarrow 0$, $T$ has  a pole at $\Omega = -\Eb = -1/(2ma^2)$ as in 2D. 
From $T$, the one-hole self-energy is obtained by closing the loop
\begin{align}
\label{Sigmaapp}
\Sigma_1(\omega) =  \int d^3\k  \ T(\omega + \epsilon_\k + i0^+).
\end{align}
\subsection*{Molecule-hole feature}
We focus on large binding energies, $\Eb\gg \mu \Leftrightarrow k_F a \ll 1$, and expand around $\omega = -\Eb$. Keeping terms up to order $\mathcal{O}(\mu^{3/2}/\Eb)$, we find (compare also \cite{combescot2007normal}): 
\begin{align}
\label{Sigma13d}
\Sigma_1(\omega) = \frac{2\sqrt{\Eb}}{\pi}\int_0^\mu d\epsilon \frac{\sqrt{\epsilon}}{\omega + \Eb + \epsilon - \frac{2}{3} \gamma\mu + i0^+}, 
\end{align} 
with $\gamma = k_Fa/\pi$ as defined in Eq.\ \eqref{gammadef} of the main text. From Eq.\ \eqref{Sigma13d} one can deduce, reattaching the hole lines
\begin{align}
&H_1(\omega) \simeq \\ & \frac{2\sqrt{\mu}}{\pi\Eb^{3/2}} \ln \left(\frac{\omega + \Eb - \omega_0 + i0^+}{-\mu}\right), \  \omega_0 =  - \mu + \tfrac{2}{3}\gamma \mu,  \notag 
\end{align} 
which holds for $|\omega + \Eb - \omega_0| \ll \mu$. This result is very similar to 2D, Eq.\ \eqref{H12D}, apart from a different non-universal prefactor, and a shift of the molecule feature by $\omega_0$. Thereby, already at one-hole level the molecule is placed at the right energy $\omega \simeq - \Eb + \omega_0 \simeq \omega_{\text{th},1}$ up to order $\mathcal{O}(a)$, as can be checked by inserting Eq.\ \eqref{phaseshift3d_eq} into \eqref{Fumithreshold}. The energy shift might cause some technical modifications at the two-hole level, that is, for diagrams $H_2$ in Fig.\ \ref{diagfig} of the main text (whose 3D evaluation is beyond the scope of this work), but the overall 2D strategy should remain valid. 
Let us note that resumming the self-energy of Eq.\ \eqref{Sigma13d} with Dyson's equation (which is incorrect for large masses as explained in the main text) yields a spurious attractive polaron, determined from 
\begin{align}
\omega - \Sigma_1(\omega) = 0.
\end{align}
For $\Eb \gg \mu$ this equation is readily solved, and yields $\omega = \omega_{\text{th,1}} - \mathcal{O}\left(\mu\exp(-1/\gamma)\right)$, i.e.\ the gap between the continuum and the polaron is only exponentially small. Correct evalution of $H_2$ and higher diagrams should eliminate the polaron as in 2D. 

\subsection*{Repulsive polaron}
For $\omega \gtrsim 0$, $\Sigma_1(\omega)$ reads
\begin{align} \notag 
&\Sigma_1(\omega)  = \\   &\int_0^\mu d\epsilon \frac{\sqrt{\epsilon}}{\sqrt{\mu}/{\gamma} - \left(2\sqrt{\mu} + \sqrt{\omega + \epsilon} \ln \left|\frac{\sqrt{\mu} - \sqrt{\omega + \epsilon}}{\sqrt{\mu}+ \sqrt{\omega + \epsilon}}\right|\right)}. 
\end{align}
We restrict ourselves to frequencies $\mu \exp(-1/\gamma) \ll \omega \ll \mu$ similar to 2D (except that the small parameter $\gamma$ is not logarithmic anymore). Then we find, with l.a.: 
\begin{align}
\Sigma_1(\omega) = \frac{2}{3} \gamma \mu +  \gamma^2 \omega \ln\left(\frac{\omega}{\mu}\right) - i\pi \gamma^2 \omega \theta(\omega), 
\end{align}
which is in full agreement with the 2D result of Eq.\ \eqref{sigmaexp} except for the factor $2/3$  dictated by Fumi's theorem. 

\section{$M = \infty$: Evaluation of $H_2$ in the time domain in 2D}
\label{timedom_supp}

\begin{figure}
\centering
\includegraphics[width=.6\columnwidth]{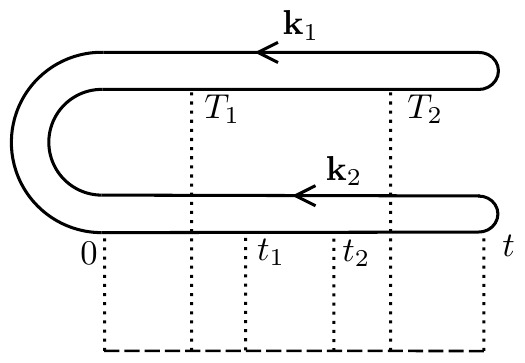}\caption{A representative of the series $\Sigma_2^a$, see also Fig.\ \ref{diagfig} of the main text.}
\label{Sigma2afig}
\end{figure}

The evaluation of the diagrams $H_2$ drawn in Fig.\ \ref{diagfig} is similar to Ref.\ {\cite{pimenov2017fermi}}. Let us first focus on the diagrammatic series $H_2^a$, and evaluate the corresponding contribution to the self-energy part $\Sigma_2^a$, i.e., amputate the external impurity lines first. The relevant diagram is redrawn in Fig.\ \ref{Sigma2afig}.
 We specialize on energies $\omega \simeq -\Eb$.
When the energy is measured from the impurity level, the time-domain impurity Green function for infinite mass reduces to a step-function: $D_0(t) = -i\theta(t)$. Thus, the impurity lines impose the time-ordering of the interactions only. We parenthetically note that for finite mass, the impurity propagator aquires a non-trivial momentum-dependence, which obstructs the time-domain evaluation, and is the reason for going into the more complicated frequency domain calculation in the next Section. 
The general expression for all diagrams which preserve the structure of $\Sigma_2^a$, with the interaction lines at initial and final times connecting to the lower part of the ``horseshoe", reads

\begin{widetext}
\begin{align}
\label{H2afirst}
 \Sigma_2^a(t) =
&-i V_0^2 \sum_{n=1}^\infty (-V_0)^n \sum_{m = 0}^\infty (-V_0)^m  \theta(t) \int  d\k_x d\k_y \\  & \notag
\int_{0<T_1 \hdots < T_n <t} dT_1 \hdots dT_n\int d\q_1 \hdots d\q_{n-1} G(\k_x, T_1 -t) G(\q_2, T_2 - T_1) \hdots G(\q_{n-1}, T_n - T_{n-1})  G(\k_y, - T_n) \\
\notag &\int_{0<t_1\hdots< t_m<t} dt_1 \hdots dt_m  \int d\p_1 \hdots \int d\p_{m+1} G(\p_1, t_1) \hdots G(\p_{m},t_m - t_{m-1}) G(\p_{m+1}, t-t_m) ,
\end{align}
where $G(\k, t) = -i(\theta(t) - n_\k) \exp(-i\epsilon_\k t)$, and $n_\k = \theta(k_F - k)$ is the zero temperature Fermi function. Introducing retarded Green functions as $G^R(t)  = G(t) \theta(t)$, Eq.\ \eqref{H2afirst} can be rewritten as:
\begin{align}
 &\Sigma_2^a(t) = -i V_0^2 \int_{< k_F} d\k_x d\k_y \exp(i(\epsilon_{\k_x} + \epsilon_{\k_y})t)\cdot A(t) B(t) \\
&A(t) = \sum_{n = 1}^\infty (-V_0)^n \int d\q_1\hdots d\q_{n-1} \left[  G^R(\k_x, \cdot) *  G^R(\q_1, \cdot)  * \hdots *  G^R(\q_{n-1}, \cdot) *  G^R(\k_y, \cdot)\right](t) \\
&B(t)= \sum_{m = 0}^\infty (-V_0)^m \int d\p_1 \hdots d\p_{m+1} \left[  G^R(\p_1, \cdot) * \hdots *  G^R(\p_{m+1}, \cdot)\right](t), 
\end{align}
where $*$ denotes convolutions, $[f*g](t) = \int\! d\tilde t f(t - \tilde t) g(t)$. 
\end{widetext}

\begin{figure}
\centering
\includegraphics[width=.9\columnwidth]{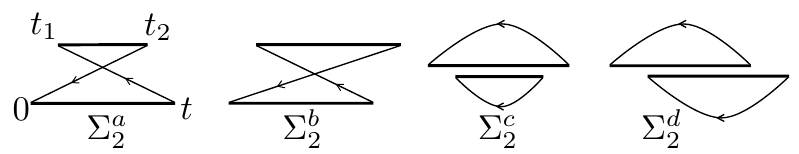}
\caption{Two-hole diagrams $\Sigma_2$ (with amputated impurity lines) in bold-line representation. Initial, final and intermediated times are indicated for $\Sigma_2^a$ only.}
\label{H2bold}
\end{figure}

Fourier-transformation using the convolution theorem turns the convolutions into a geometric series, which are resummed in the same way as  the $T$-matrix, Eq.\ \eqref{Tmatrixdef}. Changing to energy integrations for the remaining two momentum integrals over $\k_x$ and $\k_y$ results in
\begin{align}
\notag
&\Sigma_2^a(\omega) = -i \int_0^\mu dx dy \int_{-\infty}^\infty \frac{d\omega_1}{2\pi} \frac{1}{\omega_1 -x + i0^+} \frac{1}{\omega_1 - y + i0^+}  \\ &\frac{1}{\ln\left(\frac{\omega_1 - \mu + i0^+}{-\Eb}\right)} \frac{1}{\ln\left(\frac{\Omega_1 - \mu +i0^+}{-\Eb}\right)} , \quad \Omega_1 = \omega  + x + y - \omega_1. \label{H2awithsigns}
\end{align}
For the remaining $\omega_1$-integration we use the following approach: we split the $1/\ln$-terms into a part containing a pole and a part containing a branch cut:
\begin{align} \label{polebranchsplit}
&\frac{1}{\ln\left(\frac{\omega_1 - \mu + i0^+}{-\Eb}\right)} = \frac{-\Eb}{\omega_1 + \Eb - \mu + i0^+} \\ & + \left(\frac{1}{\ln\left(\frac{\omega_1 - \mu + i0^+}{-\Eb}\right)} + \frac{\Eb}{\omega_1 + \Eb - \mu + i0^+}\right). \notag \end{align}
The first part, containing the pole, can be interpreted as bound state propagator, while the second part, containing the branch cut, corresponds to the continuum contribution; this is also in agreement with the evaluation of $H_1$ in Sec.\ \ref{H1app}. Employing the spectral representation, it is easily shown that the combination of the branch cut contributions for both $1/\ln$-functions yields a result with vanishing imaginary part for $\omega \simeq -\Eb$. Since the spectrum is determined by the latter, we omit this part. The remainder is evaluated using Cauchy's theorem:
\begin{align}
 \label{Sigma2adom}
&\Sigma_2^a(\omega) =  \int_0^\mu \!dx dy \frac{\Eb}{\omega + \Eb + x - \mu + i0^+} \\ \notag &\frac{1}{\omega + \Eb + y - \mu + i0^+} \frac{1}{\ln\left(\frac{\omega + \Eb + x + y - 2\mu + i0^+}{-\Eb}\right)}.
\end{align}
Restricting to $\mu^2/\Eb \ll \omega + \Eb \ll \mu$ as explained below Eq.\ \eqref{sigmaexp}, we find, in agreement with Eq.\  \eqref{H1+H2}.
\begin{align}
\label{H2aappfinal}
H_2^a( \omega) \simeq - \frac{\gamma}{\Eb} \ln^2\left(\frac{\omega + \Eb + i0^+}{-\mu}\right).
\end{align}
The diagrams $H_2^{c,d}$ can be evaluated along the same lines; the evaluation of $H_2^b$ requires a ``generalized convolution theorem":
\begin{align}
\label{genconv}
&\mathcal{F}\left(\int_{-\infty}^{\infty} dt_1 f(t-t_1)  g(t,t_1)\right)(\Omega) = \\ &\int_{-\infty}^{\infty} \frac{d\omega_1}{2\pi} f(\omega_1) g(\Omega - \omega_1, \omega_1), \notag
\end{align}
where $\mathcal{F}$ denotes the Fourier transform, $\mathcal{F}(f)(\Omega) = \int dt f(t) \exp(i\Omega t)$,  and $f$ and $g$ are any two well-behaved functions; for more details, see Appendix B.2 of Ref.\  \cite{pimenov2017fermi}. There is, however, a quicker way to arrive at the results: one can combine the impurity interacting with a forward propagating electron into a ``bold" propagator of the form
\begin{align}
 \quad D_\text{b}(t) \propto  \int_{-\infty}^{\infty} \frac{d\omega_1}{2\pi} \exp(-it\omega_1) \frac{\theta(t)}{\ln\left(\frac{\omega_1 - \mu + i0^+}{-\Eb}\right)} ,
\end{align}
The self-energy parts of the series $H_2$ of Fig.\ \ref{diagfig} can be redrawn in this bold-line representation as shown in Fig.\ \ref{H2bold}. The thin lines correspond to the two holes, with Green function $G(x,t) = i\theta(-t)  n_F(x) \exp(-ixt)$, where $x$ is an energy variable as in Eq.\ \eqref{H2awithsigns}.
With these ingredients, $ \Sigma_2^a(t)$ can immediately be written down in a closed form up to an overall phase factor:
\begin{align}\notag
& \Sigma_2^a(t) \propto \int_0^\mu\! dx dy \int_0^t \!  dt_2 \int_0^{t_2} \!  dt_1 \\ & D_\text{b}(t_2 - t_1) D_\text{b}(t) G(x,-t_2) G(y,t_1 - t)
\end{align}
By Fourier transformation one easily reproduces Eq.\ \eqref{H2awithsigns}; the overall phase is fixed referring to the ordinary diagrams.  The time domain representations of the remaining diagram read
\begin{align}
\label{H2btime}
&\Sigma_2^b(t) \propto \int_0^\mu dx dy \int_0^t dt_2 \int_0^{t_2} dt_1 \\ & D_\text{b}(t-t_1) D_\text{b}(t_2) G(x,-t) G(y,t_1 - t_2), \notag
\\
&
\Sigma_2^c(t) \propto \int_0^\mu dx dy \int_0^t dt_2 \int_0^{t_2} dt_1 \\   &D_\text{b}(t) G(x,-t) D_\text{b}(t_2 - t_1) G(y,t_1 - t_2), \notag \\  &
\Sigma_2^d (t) \propto \int_0^\mu dx dy \int_0^t dt_1 \int_0^t dt_2 \\ & D_\text{b}(t_1) G(x,-t_1)  D_\text{b}(t-t_2) G(y,t_2 - t). \notag
\end{align}
Fourier transforming with help of Eq.\ \eqref{genconv}, and splitting into pole and branch cut contributions as described in Eq.\ \eqref{polebranchsplit}, we find for $H_2^{b,c}$\begin{align}\notag
&H_2^b(\omega) + H_2^c(\omega) \simeq \frac{1}{\Eb}  \int_0^\mu dx
\frac{1}{(\omega + \Eb + x - \mu + i0^+)^2} \\ & \cdot \left(-\mu - \int_0^\mu dy \frac{1}{\ln\left(\frac{\omega + \Eb + x + y - 2\mu + i0^+}{-\Eb}\right)}\right).
\label{H2bctimedom}
\end{align}
Comparison with Eqs.\ \eqref{Sigmaapp}, \eqref{sigmaexp} shows that these contributions shift the molecule threshold, $\Eb \rightarrow \Eb + \mu(1-\gamma)$, as claimed in the main text. This shift is the only relevant self-energy effect: the non-trivial frequency-dependence of the molecule self-energy involves a factor $\gamma^2 \ln[(\omega + \Eb)/(-\mu)]$, which is  subleading compared to Eq.\ \eqref{H2aappfinal}.

In addition, we find that $H_2^d$ is subleading with an extra factor $\mu/\Eb$ compared to the other diagrams, and we may therefore savely neglect it.

\section{$M = \infty$: Evaluation of $H_2$ in the frequency domain in 2D}
\label{freq-domain}
The evaluation of $H_2$ in the time-domain is  instructive, but crucially depends on the fact that the impurity is dispersionless. Thus, it does not simply generalize to finite mass impurities. To circumvent this problem, we first recover the results of the previous section in the frequency domain, which allows for extension to the finite mass case.
In this approach, the diagrams are organized by the number of $T$-matrices. For $H_2^a$, the first two diagrams are shown in Fig.\ \ref{diagfig}(b). The lowest order diagram (3 $T$-matrices) reads
\begin{align}
\label{H21a_first}
&H_{2,1}^a(\omega) \simeq \\ \notag &\frac{1}{E_\text{b}^2} \! \int_0^\mu \!dx dy \int_\mu^\xi \!d\epsilon_p \frac{1}{\ln\left(\frac{\omega + x - \mu + i0^+}{-\Eb}\right)} \frac{1}{\ln\left(\frac{\omega + y - \mu + i0^+}{-\Eb}\right)} \\ & \frac{1}{\ln\left(\frac{\omega +x + y - \epsilon_p - \mu + i0^+}{-\Eb}\right)} \frac{1}{\omega - \epsilon_p + x + i0^+} \frac{1}{\omega - \epsilon_p + y + i0^+}.\notag
\end{align}
For $\omega \simeq -\Eb$, the $\epsilon_p$-integral is dominated by the pole of the third logarithm, around which we can expand. The remaining Green functions effectively cut off the integration at $\epsilon_p \simeq \Eb$. Since this scale only appears in the argument of a logarithm (see below), an exact determination is not required within l.a.. Thus, we can approximate
\begin{align}
\label{H2awithI}
&H_{2,1}^a(\omega) \simeq
-\frac{1}{\Eb}\! \int_0^\mu \!dx dy \\ &\frac{1}{\omega + \Eb + x - \mu + i0^+} \frac{1}{\omega + \Eb + y - \mu + i0^+} \cdot  I \notag \\
\label{Idef}
&I = \int_\mu^\Eb d\epsilon_p \frac{1}{\omega + \Eb + x + y - \epsilon_p - \mu + i0^+} \simeq \\  &\ln\left(\frac{\omega + \Eb + x + y - 2\mu + i0^+}{-\Eb}\right), \ \omega \simeq -\Eb. \notag
\end{align}
A similar evaluation of the second diagram of the series $H_2^a$ reproduces Eq.\ \eqref{H2awithI} with $I$ replaced by $I^3$. Extrapolating this behavior, the higher order diagrams  yield a series $I + I^3 + I^5 + \hdots = I/(1-I^2) \simeq 1/(-I)$. In total:
\begin{align}
 \label{H2adom}
&H_2^a(\omega) =  \frac{1}{\Eb} \int_0^\mu \!dx dy \frac{1}{\omega + \Eb + x - \mu + i0^+} \\ \notag &\frac{1}{\omega + \Eb + y - \mu + i0^+} \frac{1}{\ln\left(\frac{\omega + \Eb + x + y - 2\mu + i0^+}{-\Eb}\right)},
\end{align}
in agreement with Eq.\ \eqref{Sigma2adom}. The remaining expressions $H_2^{b,c,d}$ can be evaluated along the same lines. For $H_2^{b,c}$, the result is in perfect agreement with Eq.\ \eqref{H2bctimedom}. For $H_2^d$, one arrives at
\begin{align}
&H_{2}^d(\omega) =  \frac{1}{\Eb} \int_0^\mu dx dy \\ &\frac{1}{\omega + \Eb + x - \mu + i0^+} \frac{1}{\omega + \Eb + y - \mu + i0^+} \cdot \frac{1}{I^2}. \notag
\end{align}
Since $I$ is a large logarithm of order $1/\gamma$, this result is subleading, although only by a factor $\gamma$ and not $\mu/\Eb$ as in the time domain; this slight discrepancy can  be attributed to the evaluation with l.a.

\section{The molecule continuum for finite mass}
\label{mass-app}
With the infinite mass results for $H_1, H_2$ at hand, we  move to the finite mass case, focusing on $\omega \simeq -\Eb$. We start from re-evaluation of Eq.\ \eqref{Tmatrixdef} to leading order in $\beta = m/M$:
\begin{align}
T(\omega + \epsilon,\k) = - \frac{1+\beta}{\rho} \frac{1}{\ln\left(\frac{\omega + \epsilon - \mu\beta - \mu - k^2/{2M_+} + i0^+}{-\Eb}\right)},
\end{align}
where $M_+ = m + M$, and the vacuum binding energy $\Eb$ now reads $(1+\beta)\xi e^{-(1+\beta)/(\rho V_0)}$. In the following, we will only resolve the factors $(1+\beta)$ in the numerator of the logarithms, since the other factors just rescale the energies. Closing the contour to obtain $H_1(\omega)$, we find
\begin{align}
H_1(\omega) &\simeq \frac{1}{\Eb} \int_0^\mu dx
\frac{1}{\omega + \Eb - \beta\mu + x - \mu - \beta x + i0^+} \notag \\& \simeq
\ln\left(\frac{\omega + \Eb - 2\beta\mu + i0^+}{-\mu}\right).
\label{H1masslog}
\end{align}
The resulting contribution to the spectrum is given by
\begin{align}
\label{energycons}
&-\text{Im}\left[H_1(\omega)\right] \\ &= \frac{1}{\Eb} \int_0^\mu dx \delta\left(\omega - (-\Eb + \beta\mu  + \beta x + (\mu - x))\right) \notag.
\end{align}
The energy-conservation imposed by the delta-function describes the following process: an impurity with energy $\omega$ decays into a bound state, with ``potential energy" $-\Eb + \beta\mu$ and kinetic energy $\beta x$, and a hole with energy $(\mu - x)$. At the treshold, the hole peels off right at the Fermi surface, and the bound state has a kinetic energy $\beta\mu$.
The modification of the binding energy in the presence of a Fermi sea $-\Eb \rightarrow -\Eb + \beta\mu$ is only of secondary importance, since there are further molecule self-energy diagrams with renormalize the binding energy anyway (see Eq.\ \eqref{H2bctimedom}). We will not evaluate these in detail for finite mass, and just write $\nu_1 = \omega + \Eb - \beta\mu$ henceforth. Thus, we see that the logarithm in Eq.\ \eqref{H1masslog} is peaked at the ``direct threshold'' $\nu_1 = \beta\mu$, involving the creation of a molecule with momentum $k_F$.

Next, we evaluate the lowest order contribution (3 $T$-matrices) to diagram $H_2^a$. Resumming the logarithms, the finite mass generalization of Eq.\ \eqref{H21a_first} reads
\begin{widetext}
\begin{align}
&H_{2,1}^a(\omega) = \frac{1}{E_\text{b}^2\rho^3} \int d\k_x  d\k_y \int d\p \frac{1}{\omega - \epsilon_p + x  - E_{\k_x- \p} + i0^+} \frac{1}{\omega - \epsilon_p + y - E_{\k_y - \p} + i0^+}  \\ \notag & \frac{1}{\ln\left(\frac{\omega + x - \mu(1+\beta) - E_{\k_x} + i0^+}{-\Eb}\right)} \frac{1}{\ln\left(\frac{\omega + y - \mu(1+\beta) - E_{\k_y} + i0^+}{-\Eb}\right)} \frac{1}{ \ln\left(\frac{\omega + x + y- \epsilon_p - \mu(1+\beta) - E_{\k_x + \k_y - \p} + i0^+}{-\Eb}\right)}.
\end{align}
\end{widetext}
To compute the $\p$-integral, we expand the last $1/\ln$-function around its pole. The resulting non-trivial logarithmic integral reads
\begin{align}
\label{Itildedef}
&\tilde I = \frac{1}{\rho} \int_{k_F< \p < \sqrt{2m\Eb}} d\p \\& \frac{1}{\nu_1 + x + y -\mu -  \epsilon_p  - E_{\k_x + \k_y - \p} + i0^+}, \notag
\end{align}
c.f.\ Eq.\ \eqref{Idef}. Integration with logarithmic accuracy (which only gives access to Re$[\tilde I]$) yields
\begin{align}
\text{Re} [\tilde I] \simeq \ln\left(\frac{\text{max}(\nu_1 + x + y - 2\mu, \beta\mu)}{-\Eb}\right).
\end{align}
Postponing evaluation of $\text{Im}[\tilde I]$, one can extrapolate to the full series $H^a_{2,1}$ as in Sec.\ \ref{freq-domain}. Repeating this procedure for $H_2^{b,c,d}$, we find
\begin{align}
\notag
&H_{2}^a(\nu_1) \simeq \phantom{+}\frac{1}{\Eb\rho^2}\! \int\! d\k_x d\k_y \frac{1}{\nu_1 + x - \mu - E_{\k_x} +i0^+} \\ &\frac{1}{\nu_1 + y - \mu - E_{\k_y} + i0^+} \cdot \frac{1}{\tilde I}  \label{H2a_finitemass} \\
\notag
&H_{2}^b(\nu_1) \simeq -\frac{1}{\Eb\rho^2}  \!\int \! d\k_x d\k_y \frac{1}{(\nu_1 + x - \mu - E_{\k_x} + i0^+)^2} \\ & \cdot \left(1 + \mathcal{O}\left(1/\tilde I^2\right)\right) \label{H2b_finitemass}
\\
\label{H2c_finitemass}
&H_{2}^c(\nu_1) \simeq -\frac{1}{\Eb\rho^2}  \!\int \!d\k_x d\k_y \frac{1}{(\nu_1 + x - \mu - E_{\k_x} + i0^+)^2}\cdot\frac{1}{\tilde I} \\\
&
\notag
H_{2}^d(\nu_1) \simeq +\frac{1}{\Eb\rho^2}  \int d\k_x d\k_y \frac{1}{(\nu_1 + x - \mu - E_{\k_x} + i0^+)} \\ &\frac{1}{(\nu_1 + y - \mu - E_{\k_y} + i0^+)}\cdot\frac{1}{\tilde I^2}
\label{H2d_finitemass}.
\end{align}
Since $\tilde I$ is still a large logarithm of order $1/\gamma$, $H_2^d$ is subleading as for infinite mass. The other contributions behave as follows: with l.a., $H_2^a$ reads \begin{align}
\label{H2anu1}
H_2^a(\nu_1) \simeq - \frac{\gamma}{\Eb} \ln^2\left[ (\nu_1 - \beta\mu + i0^+)/(-\mu)\right],
\end{align}
i.e., essentially the same result as in the infinite mass case, Eq.\ \eqref{H2aappfinal}, except that the peak of the logarithm is at the direct threshold, as discussed below Eq.\ \eqref{energycons}. $H_2^{b,c}$ again act as molecular self-energy terms. First, their real parts lead to a shift of $\Eb$, which we do not compute. More importantly, the imaginary part of $H_2^c$ cuts off the logarithmic singularity at the direct threshold. This can be seen extracting the molecule self-energy part from Eq.\  \eqref{H2c_finitemass}:
\begin{align}
\Sigma_{\text{mol}}(\nu_1, \k_x) = - \frac{1}{\rho} \int d\k_y \frac{1}{\tilde{I}} \ .
\end{align}
Using Eq.\ \eqref{Itildedef}, we find
\begin{align}
&\text{Im}\left[\Sigma_\text{mol}\right](\nu_1, \k_x)  =  \int d\k_y \frac{\text{Im}[\tilde I]}{\text{Re}[\tilde I]^2 + \text{Im}[\tilde I]^2} \simeq \\ & \notag-  \gamma^2 \pi \int d\k_y d\p \   \delta(\nu_1 + x + y - \mu - \epsilon_p -  E_{\k_x + \k_y - \p}).
\end{align}
Evaluation of this standard phase space integral (see e.g.\ Appendix E of Ref.\ \cite{pimenov2017fermi} for examples) for $k_x = k_F$ yields
\begin{align}\text{Im}[\Sigma_{\text{mol}}(\nu_1,k_F)] \propto -\gamma^2 \frac{\nu_1^2}{\beta\mu}.
\end{align}
Near the direct threshold, $ \nu_1 \simeq \beta\mu$, this leads to a molecule decay rate $\Gamma \propto \gamma^2 \beta\mu$. Appropriately resummed, this rate cuts all logarithms; e.g., the one-hole result of Eq. \eqref{H1masslog} is modified as
\begin{align}
\notag
 H_1(\nu_1) & \simeq \frac{1}{\Eb} \ln\left(\frac{\nu_1  - \beta\mu + i\Gamma}{-\mu}\right) \\ &\overset{l.a.}{\simeq} \frac{1}{\Eb}  \ln \left(\max[{\nu_1 - \beta\mu},\gamma^2 \beta\mu]/\mu\right),
 \label{cutlogapp}
 \end{align}
and likewise for the term $H_2^a$ in Eq.\  \eqref{H2anu1}. The physical reason for this cut-off is the decay of the molecule at the direct threshold with $k = k_F$ into a zero momentum molecule, two holes and an electron. This process shifts the threshold to the ``indirect'' one at $\nu_1 = 0$. For $0< \nu_1 \ll \beta\mu$, the spectrum is perturbative (i.e., no large logarithms need to be resummed), and can be obtained from $\text{Im}[H_2^a, H_2^c]$, Eqs.\  \eqref{H2a_finitemass}, \eqref{H2c_finitemass}. For spinless electrons, these contributions cancel to leading order. For spinful electrons, $H_2^c$ incurs an extra factor of two, and the perturbative spectrum reads
\begin{align}
&\mathcal{A}_{\text{pert}}(\nu_1)  \simeq  \gamma^2 \frac{1}{\Eb (\beta\mu)^2} \frac{\pi}{\rho^2} \int d\k_x d\k_y d\p \\ &\delta (\nu_1 + x + y - \epsilon_p - \mu   - E_{\k_x + \k_y - \p})
\propto \frac{\gamma^2}{\Eb} \left(\frac{\nu_1}{\beta\mu}\right)^3 \!\theta(\nu_1)
\notag
\end{align}
in 2D, while in 3D the extra phase space restriction should lead to $\mathcal{A}_{\text{pert}} \propto \nu_1^{7/2}$ \cite{PhysRevB.35.7551}. Exponentiating the cut-off logarithms \eqref{cutlogapp} with a correct imaginary part to capture the perturbative spectrum yields Eq.\ \eqref{cutoffFermiedge} of the main text; the square root is yet another, continuous reformulation of the logarithm cutoff.

\section{$M = \infty$: Repulsive polaron from the linked-cluster approach}
\label{linked cluster}
\begin{figure}
\centering
\includegraphics[width=\columnwidth]{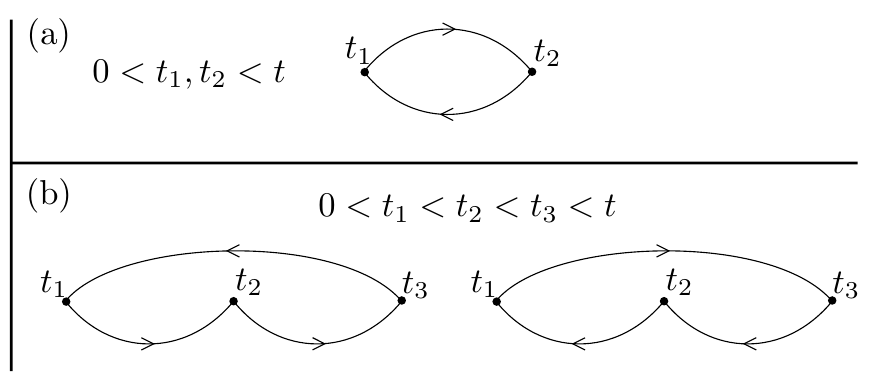}
\caption{(a) $n=2$ cluster diagram. Full lines indicate electron Green functions, and dots $V_0$-insertions. Internal times $t_1, t_2$ independently range from $0$ to $t$. (b) \emph{Time-ordered} $n=3$ cluster diagrams. The left diagram shows an ordering with one hole, which is invariant under cyclic permutation of times. The right diagram contains two holes. }
\label{clusterdiag}
\end{figure}

The leading contribution to the repulsive polaron for $M = \infty$ was already obtained in Eq.\ \eqref{H1nu2}. The full power law singularity can be reproduced in a linked cluster approach (see also Refs.\ \cite{G.D.Mah2000}, \cite{ PhysRevB.42.6850}). One starts from the following set of identities for the impurity propagator
\begin{align}
\label{clusterexpansion}
&D(t) = -i\theta(t) \braket{0|S(t)|0} \\
&\braket{0|S(t)|0} = \exp\left(\sum_n F_n(t)\right) \\
\label{Fns}
& F_n(t) = \frac{(-i)^n}{n}\!\int_0^t \!dt_1 \hdots \int_0^t\! dt_n \braket{0|\hat{T}\left\{\hat V(t_1) \hdots \hat{V}(t_n) \right \} |0} \\
&\hat{V}(t_i) = -V_0 \sum_{\k,\p} c^\dagger_\k c_\p \theta(t_i) \theta(t - t_i),
\end{align}
where $S(t)$ is the $S$-matrix, and $\hat{T}$ the time-ordering operator. Note that the impurity has effectively been eliminated from the problem, which results in Feynman diagrams such as those shown in Fig.\ \ref{clusterdiag}.
The expressions above imply an expansion in the bare interaction $V_0$. Our goal is to substitute this by an expansion in number of holes, getting rid of the $V_0$-dependence. Let us collect all one-hole diagrams: one is drawn in Fig.\ \ref{clusterdiag}(a). It is more convenient to reexpress it imposing a time-ordering $0<t_1<t_2<t$. This results in a factor of two which cancels the factor $1/2$ in Eq.\ \eqref{Fns} for $n=2$. Generalizing this approach, the one-hole diagrams can be extracted by drawing ``loop diagrams'' containg $n$ interaction insertions at times $t_1, \hdots t_n$, and reordering them in such  a fashion that only one electron propagates backwards in time; this cancels the factor $1/n$ in Eq.\ \eqref{Fns}. An example for $n=3$ is shown in Fig.\ \ref{clusterdiag}(b).
Performing this reorganization, at one-hole level we can write
\begin{align}
\label{Dc1}
&D(t) \simeq \exp(C(t)), \\
& \label{C1t}
C(t) = -\int \frac{d\omega}{\pi} \left(\frac{-it}{\omega} - \frac{1}{\omega^2} \left(\exp(-it\omega) -1\right)\right) N(\omega) \\ & N(\omega) = \text{Im} \left\{ \int_0^\mu dx \frac{1}{\ln\left(\frac{\omega + x - \mu + i0^+}{-\Eb}\right)} \right\}
\end{align}
where we resummed the one-hole diagrams similar to Sec.\ \ref{H1app} and employed the spectral representation of the retarded $1/\ln$-function. In Eq.\ \eqref{C1t}, the part linear in $t$ just shifts the polaron threshold, and we may omit it. $N(\omega)$ measures the phase space for scattering of polarons with Fermi electrons. Eqs.\ \eqref{Sigmaapp} and \eqref{sigmaexp} show that in the most important spectral window $\mu^2/\Eb \ll \omega \ll \mu$ we can approximate $N(\omega)  \simeq \gamma^2 \pi \omega$. Therefore, evaluation of Eqs.\ \eqref{Dc1} and \eqref{C1t} similar to Sec.\ 8.3.C of Ref.\ \cite{G.D.Mah2000} and Fourier transformation directly results in a repulsive polaron spectrum
\begin{align}
\label{repinfmass}
\mathcal{A}_2(\nu_2) &\propto \frac{\theta(\nu_2)}{\mu} \left(\frac{\nu_2}{\mu}\right)^{\gamma^2 -1}
\end{align}
with frequencies measured from the polaron threshold.

\section{The repulsive polaron for finite mass}
\label{modlinkc}

\subsection*{Modified linked-cluster approach}

The procedure above can also be adapted for finite mass (see, e.g., Sec.\ 3.6.B of \cite{G.D.Mah2000} or \cite{Pimenov2015}). In effect, we need to reevalute the phase-space factor $N(\omega)$ at one-hole level, and find, in 2D
\begin{align} 
\notag
N(\omega) &\simeq \gamma^2\pi \int d\k d\p\  \delta\left(\omega + \epsilon_\k - \epsilon_\p - E_{\p - \k}\right) \\
&\simeq \begin{cases} \gamma^2 \omega^{3/2}/{\sqrt{\beta\mu}}\theta(\omega) \ &\omega \ll \beta\mu \\ \gamma^2 \pi \omega \ &   \beta\mu \ll \omega \ll \mu,  \end{cases} 
\label{Nomegacases}
\end{align}
and $ N(\omega) \propto \omega^2  \ \text{for} \ \omega \ll \beta\mu$ in 3D. 
Thus, for energies beyond the recoil energy $\beta\mu$, the phase space factor assumes the infinite mass form. For smaller energies, the scattering phase space is suppressed, since processes where the polaron is scattered to large momenta of order $k_F$ involve a minimal energy cost of order $\beta\mu$ \cite{rosch1995heavy}.

We insert Eq.\ \eqref{Nomegacases} into \eqref{C1t} and first study the limit $t\rightarrow \infty$. Again ignoring the term linear in $t$, in the infinite mass case one can show that $C(t)$ diverges as $-\gamma^2\ln(|t|\mu)$. In contrast, for finite mass we find $\lim_{t\rightarrow \infty} C(t) \simeq \gamma^2 \ln(\beta)$. Inserted into the Green function of Eq.\ \eqref{Dc1}, this  limit gives rise to a finite quasiparticle-weight of the polaron, $Z \propto \beta^{\gamma^2}$. Again, the emergence of this quasi-particle weight is a consequence of the restricted low-energy scattering phase space, which partially reduces the repulsive polaron to its non-interacting form.
Moreover, we can extract the incoherent polaron spectrum for small detuning from the threshold $0<\nu_2<\beta\mu$ simply by expanding the exponential in Eq.\ \eqref{C1t}, since there is no large logarithmic quantity to prevent it. This yields, in 2D
\begin{align}
\mathcal{A}(\nu_2) \propto \gamma^2 \frac{1}{\sqrt{\beta\mu \nu_2}},
\end{align}
while in 3D the incoherent part is approximately constant $\propto \gamma^2 /(\beta\mu)$. For $\nu_2 \gg \beta\mu$ one recovers the infinite mass behavior.  Interpolating between these two limits yields formula \eqref{A2main},
apart from the finite width of the repulsive polaron quasiparticle to be discussed below.

\subsection*{Width of the repulsive polaron}

\begin{figure}[t]
\centering
\includegraphics[width=\columnwidth]{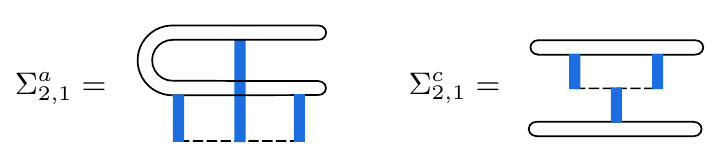}
\caption{Diagrams $\Sigma_{2,1}^{a,c}$ contributing to the decay of the repulsive polaron}
\label{H21_app}
\end{figure}

In the previous section, we only considered the decay of the single impurity into particle-hole excitations, but neglected the decay into the molecular state. To incorporate this process, we need to go to two-hole level. 
The decay rate vanishes for infinite mass, since the problem becomes single-particle, hence the molecule and repulsive polaron sectors decouple.
Indeed, starting from the exact expressions for the impurity self-energy $\Sigma_2$ in the time domain, of a form similar to Eq.\ \eqref{H2awithsigns}, one can show that $\sum_i \text{Im}[\Sigma_2^i](\omega = 0^-) = 0$, where the frequency argument $0^-$ is chosen to exclude the UV-tail arising from electron-hole excitations. 
This cancellation implies that the repulsive polaron does not acquire a Lorentzian IR-tail for infinite mass. This is not necessarily true for finite mass. An evaluation of all two-hole diagrams for $\omega= 0^-$ similar to Sec.\ \ref{freq-domain} appears too involved. A simpler estimate can be given by restriction to the ``first-order'' diagrams $\Sigma_{2,1}$ with the minimal number of $T$-matrices.
The diagrams with a  minimal number of 3 $T$-matrices and nonvanishing imaginary parts for $\omega = 0^-$ are $\Sigma_{2,1}^{a,c}$, shown in Fig.\ \ref{H21_app}.

 The rate resulting from these diagrams has been evaluated effectively for $\Eb \rightarrow 0$ in Ref.\ \cite{bruun2010decay} for 3D, but here we focus on $\Eb \gg \mu$. We approximate the central $T$-matrix by a pole to incorporate the molecule, and the remaining two $T$-matrices by $\gamma$. Taking the imaginary part, we find
\begin{align}
&\tilde \Gamma_2 \simeq \gamma^2 \frac{\Eb}{\rho^3} \! \int d\k_x d\k_y d\p \   \delta( x + y - \epsilon_p - E_{\k_x +
\k_y - \p}+ \Eb) \notag  \\& \frac{1}{ \epsilon_p - x + E_{\k_x - \p}} \left( \frac{1}{ \epsilon_p - x + E_{\k_x - \p} } - \frac{1}{ \epsilon_p - y + E_{\k_y - \p} } \right) \notag \\
&\propto \gamma^2 \frac{\mu^4}{\Eb^3}\left(1-3\beta\right),  \label{rate2eq}
\end{align}
where the last estimate holds to leading order in $\mu/\Eb, \beta$. In \eqref{rate2eq} we may omit the $\beta$-independent part, since the infinite mass cancels upon complete evaluation as discussed above. The remainder can be used to estimate
$\Gamma_2  \sim \beta \gamma^2 \frac{\mu^4}{\Eb^3}$. For 3D, we expect a similar behavior, although details of the scaling could be different.

\end{document}